\documentstyle[12pt]{article}

\begin{document}

\setcounter{page}{1}

\pagestyle{plain} \vspace{1cm}
\begin{center}
\Large{\bf Holographic Dark Energy from a Modified GBIG Scenario  }\\
\small \vspace{1cm} {\bf Kourosh Nozari$^{a,b,*}$} \quad and \quad {\bf Narges Rashidi$^{a,\dag}$}\\
\vspace{0.5cm} {\it $^{a}$Department of Physics, Faculty of Basic
Sciences,\\
University of Mazandaran,\\
P. O. Box 47416-95447, Babolsar, IRAN\\
\vspace{0.5cm} $^{b}$Research Institute for Astronomy and
Astrophysics of Maragha,
\\P. O. Box 55134-441, Maragha, IRAN\\ \vspace{0.5cm}
$^{*}$knozari@umz.ac.ir\\
$^{\dag}$ n.rashidi@umz.ac.ir}

\end{center}
\vspace{1.5cm}
\begin{abstract}
We construct a holographic dark energy model in a braneworld setup
that gravity is induced on the brane embedded in a bulk with
Gauss-Bonnet curvature term. We include possible modification of the
induced gravity and its coupling with a canonical scalar field on
the brane. Through a perturbational approach to calculate the
effective gravitation constant on the brane, we examine the
outcome of this model as a candidate for holographic dark energy.
\end{abstract}
\newpage
\section{Introduction}
It is observationally confirmed from supernovae distance-redshift
data, the microwave background radiation, the large scale structure,
weak lensing and baryon oscillations that the current expansion of
the universe is accelerating [1]. There are several approaches to
explain this late-time accelerated expansion of the universe. One of
these approaches, is to introduce some sort of unknown energy
component ( the dark energy) which has negative pressure (See [2]
and references therein). The simplest candidate in this respect is a
cosmological constant in the framework of the general relativity.
However, huge amount of fine-tuning, no dynamical behavior and also
unknown origin of emergence make its unfavorable [3]. Another
alternative is a dynamical dark energy: the cosmological constant
puzzles may be better interpreted by assuming that the vacuum energy
is canceled to exactly zero by some unknown mechanism and
introducing a dark energy component with a dynamically variable
equation of state [2]. Nevertheless, the main problem with the dark
energy is that its nature and cosmological origin are still obscure
at present. Other alternatives to accommodate present accelerated
expansion of the universe are modified gravity [4] and some
braneworld scenarios such as the Dvali-Gabadadze-Porrati (DGP)
scenario [5].  Here, we are interested in to probe the nature of
dark energy in the context of holographic dark energy models. The
idea of holographic dark energy comes from quantum gravity
considerations [6]. This model proposes that if $\rho_{\Lambda}$ is
the quantum zero-point energy density caused by a short distance
cut-off, the total energy in a region of size $L$ should not exceed
the mass of a black hole of the same size. The holographic dark
energy density corresponds to a dynamical cosmological constant and
in this respect, the standard General Relativity should be modified
by some gravitational terms which became relevant at present
accelerating universe. Modified gravity provides the natural
gravitational alternative for dark energy. Moreover, modified
gravity presents a natural unification of the early time inflation
and late-time acceleration due to different role of gravitational
terms relevant at small and at large curvature [4,7]. Also modified
gravity may naturally describe the transition from non-phantom phase
to phantom one without necessity to introduce the exotic matter.
$f(R)$ gravity is one successful attempt in this direction. Another
theory proposed as gravitational dark energy is the
scalar-Gauss-Bonnet gravity which is closely related with low-order
string effective action [8]. The possibility to extend such
consideration to third order (curvature cubic) terms in low-energy
string effective action exists too [9]. Moreover, one can develop
the reconstruction method for such theories [10]. It has been
demonstrated that some scalar-Gauss-Bonnet gravities may be
compatible with the known classical history of the universe
expansion [7].

In this paper we propose a unified $f(R)$-Gauss-Bonnet gravity with
non-minimal coupling of the scalar field to $f(R)$. We proceed a
holographic dark energy approach to examine cosmological dynamics in
this setup and we show that this model accounts for phantom-like
behavior. This model presents a smooth crossing of the phantom
divide line by its equation of state parameter and this crossing
occurs in the same way as is supported observationally, that is,
from above -1 to its below.

\section{The Setup}
The action of our modified GBIG model contains the Gauss-Bonnet term
in the bulk and modified induced gravity term on the brane where a
scalar field non-minimally coupled to induced gravity is present on
the brane
$$ S=\frac{1}{2\kappa_{5}^{2}}\int d^{5}x\sqrt{-g^{(5)}}\bigg\{R^{(5)}-
2\Lambda_{5}+\beta\bigg[R^{(5)2}-4R_{ab}^{(5)}R^{(5)ab}+R_{abcd}^{(5)}R^{(5)abcd}\bigg]\bigg\}$$
\begin{equation}
+\int_{brane} d^{4} x
\sqrt{-q}\bigg[M_{5}^{3}K^{\pm}+\frac{M_{4}^{2}}{2}\,\alpha(\varphi)f(R)-\lambda-[\bigtriangledown
\phi]^2-2V(\phi)+L_{m}\bigg]\ ,
\end{equation}
where $\beta$($\geq 0$) is the GB coupling constant,
\,$\kappa_{5}^{2}$\,is the five dimensional gravitational constant,
\,$\lambda$\,is the brane tension,  $\Lambda_{5}$ is the bulk
cosmological constant and \,$\overline{ K }$\, is the trace of the
mean extrinsic curvature of the brane.

To obtain cosmological dynamics in this setup, we use the following
line-element
\begin{equation}
ds^{2}=q_{\mu\nu}dx^{\mu}dx^{\nu}+b^{2}(y,t)dy^{2}=-n^{2}(y,t)dt^{2}+
a^{2}(y,t)\gamma_{ij}dx^{i}dx^{j}+b^{2}(y,t)dy^{2}\,,
\end{equation}
where $\gamma_{ij}$ is a maximally symmetric 3-dimensional metric
defined as $\gamma_{ij}=\delta_{ij}+k\frac{x_{i}x_{j}}{1-kr^{2}}$
and $k=-1,0,1$ parameterizes the spatial curvature. The generalized
cosmological dynamics of this setup is given by the following
Friedmann equation ( see [11] for details)
$$ \bigg[1+\frac{8}{3}\beta \bigg(H^{2}+\frac{\Phi}{2}+\frac{K}{a^{2}}\bigg)\bigg]^{2}
\bigg(H^{2}-\Phi+\frac{K}{a^{2}}\bigg)=\Bigg\{r\alpha(\varphi)\bigg[\bigg(H^{2}+\frac{K}{a^{2}}\bigg)f'(R)\bigg]$$
\begin{equation}
-\frac{\kappa_{5}^{2}}{6}\bigg[\rho_{m}+\rho_{\varphi}+\lambda+\frac{M_{4}^{2}}{2}\,\alpha(\varphi)\bigg(Rf'(R)-f(R)
-6H\dot{R}f''(R)\bigg)\bigg]\Bigg\}^{2}\ ,
\end{equation}
where a dash denotes $\frac{d}{dR}$ and a dot marks $\frac{d}{dt}$.
The energy-density corresponding to the non-minimally coupled scalar
field is as follows
\begin{equation}
\rho_{\varphi}=\bigg[\frac{1}{2}\dot{\phi}^{2}+n^{2}V(\varphi)-
6f(R)\frac{d\alpha(\varphi)}{d\varphi}H\dot{\varphi}\bigg]_{y=0},
\end{equation}
where $y$ is the coordinate of the fifth dimension and the brane is
located at $y=0$. To proceed further, we set $n^{2}(y,t)|_{y=0}=1$.
Now we solve analytically the friedmann equation (3). It is
convenient to introduce the dimensionless variables ( see for
instance [12])
\begin{equation}
\bar{H}=\frac{8}{3}\frac{\beta }{r
\alpha(\varphi)f(R)}H=2\frac{\Omega_{\beta}\sqrt{\Omega_{r}}}{\alpha(\varphi)f(R)}E(z)\,,
\end{equation}
\begin{equation}
\bar{\rho}=\frac{32}{27}\frac{\kappa_{5}^{2}\beta^{2}}{r^3
 \alpha^{3}(\varphi)f^{3}(R)}\Big(\rho_{m}+\alpha(\varphi)\rho_{\varphi}+\rho_{c}\Big)=
 4\frac{\Omega_{r} \Omega_{\beta}^{2}}{\alpha^{3}(\varphi)f^{3}(R)}
 \Big[\Omega_{\varphi}+\alpha(\varphi)\Omega_{c}+\Omega_{m}(1+z)^{3}\Big]\,,
\end{equation}
and
\begin{equation}
b=\frac{8}{3}\frac{\beta}{r^2
\alpha^{2}(\varphi)f^{2}(R)}=4\frac{\Omega_{\beta}
\Omega_{r}}{\alpha^{2}(\varphi)f^{2}(R)}\,,
\end{equation}
where $r\equiv \frac{\kappa^{2}_{5}}{2\kappa^{2}_{4}}$ is the DGP
crossover scale and by definition
\begin{equation}
E(z)\equiv\frac{H}{H_{0}}\,,\,\Omega_{m}\equiv\frac{\kappa_{4}^{2}\rho_{m0}}{3H_{0}^{2}}\,,\,\,\Omega_{\varphi}
\equiv\frac{\kappa_{4}^{2}\rho_{\varphi}}{3H_{0}^{2}}\,,\,\,\Omega_{r}\equiv\frac{1}{4r^{2}H_{0}^{2}}\,,\,\,\Omega_{\beta}
\equiv\frac{8}{3}\beta H_{0}^{2}\,,\,
\Omega_{c}\equiv\frac{\kappa_{4}^{2}\rho_{c}}{3H_{0}^{2}}\,,
\end{equation}
where
$$\rho_{c}\equiv \frac{M_{4}^{2}}{2}\bigg(Rf'(R)-f(R)
-6H\dot{R}f''(R)\bigg).$$ The Friedmann equation (3) now takes the
following compact form
\begin{equation}
\bar{H}^{3}+\bar{H}^{2}+b\bar{H}-\bar{\rho}=0.
\end{equation}
The number of real roots of this equation is determined by the sign
of the discriminant function N defined as [13]
\begin{equation}
N=Q^{3}+R^{2}
\end{equation}
where
\begin{equation}
Q\equiv\frac{1}{3}(b-\frac{1}{3})
\end{equation}
\begin{equation}
R\equiv\frac{1}{6}b+\frac{1}{2}\bar{\rho}-\frac{1}{27}\,\, .
\end{equation}
If $N>0$,  then there is a unique real solution. If $N<0$,  there
are $3$ real solutions. Finally, if $N=0$, all roots are real and at
least two are equal. In which follows, we consider the case with
$N>0$ and in this case the unique real solution is given as follows
[12,13]
\begin{equation}
\bar{H}=\frac{1}{3}\Bigg[2\sqrt{1-3b}\,\cosh(\frac{\eta}{3})-1\Bigg]
\end{equation}
where
\begin{equation}
\cosh(\eta)=\frac{R}{\sqrt{-Q}}\,\,,\quad\quad
\sinh(\eta)=\sqrt{\frac{Q^{3}+R^{2}}{-Q^{3}}}\,\,.
\end{equation}
Therefore, we achieve the following solution for the Friedmann
equation (3)
\begin{equation}
H=\frac{r\alpha(\varphi)f(R)}{8\beta}\Bigg[2\sqrt{1-3b}\,\cosh(\frac{\eta}{3})-1\Bigg].
\end{equation}

After obtaining a solution of the Friedmann equation, in the next
section we try to provide the basis of a holographic dark energy in
this setup.
\section{Effective Gravitational Constant}
Now we derive the gravitational constant within a perturbation
theory to construct our holographic dark energy scenario. Firstly,
we analyze the weak field limit of the model presented in the
previous section within the slow time-variation approximation and at
small scales with respect to the horizon's size. It is convenient to
use the longitudinal gauge metric written directly in terms of
$\zeta=\log a$
\begin{equation}
ds^{2}=e^{2\zeta}\bigg[-(1+2\Psi)\frac{d\zeta^{2}}{a^{2}H^{2}}+(1-2\Upsilon)dx_{i}dx^{i}\bigg]
\end{equation}
The field $\Psi$ is the Newtonian potential and $\Upsilon$ is the
leading-order, spatial post-Newtonian correction which will permit
us to calculate the stress-anisotropy. For the longitudinal
post-Newtonian limit to be satisfied, we require
$\Delta\Upsilon>>a^{2}H^{2}\times(\Upsilon,\dot{\Upsilon},\ddot{\Upsilon})$
and similarly for other gradient terms [14]. For a plane wave
perturbation with wavelength $\bar{\lambda}$\,, we see that
$H^{2}\Upsilon$ is much smaller than $\Delta\Upsilon$ when
$\bar{\lambda}<<\frac{1}{H}$ . The requirement that $\dot{\Upsilon}$
is also negligible implies the condition
\begin{equation}
\frac{d\log \Upsilon}{d \zeta}<<\frac{1}{(\bar{\lambda} H)^{2}}\,,
\end{equation}
which holds if the condition $\bar{\lambda}<<\frac{1}{H}$ is
satisfied for perturbation growth. This argument can be applied for
$\ddot{\Upsilon}$,\, $\Upsilon$ and $\delta \varphi$. Now using the
metric (16), we can rewrite the $(\zeta\zeta)$ and $(ij)$ components
of the gravitational field equations and the scalar field equation
as follows [14]
\begin{equation}
\frac{3}{2}\,a^{2}H^{2}\delta\Omega_{m}=A\Delta\Upsilon+B\frac{\Delta\delta\varphi}{\dot{\varphi}}
\end{equation}
\begin{equation}
A\Delta\Psi=C\Delta\Upsilon+D\frac{\Delta\delta\varphi}{\dot{\varphi}}
\end{equation}
\begin{equation}
B\Delta\Psi=D\Delta\Upsilon-E\frac{\Delta\delta\varphi}{\dot{\varphi}}
\end{equation}\\
where $H$ is given by (15) and we have defined the following
quantities
\begin{equation}
A=1+\frac{4H^{2}\beta\dot{\varphi}^{2}}{3},
\end{equation}
\begin{equation}
B=-\frac{24H^{2}\beta\dot{\varphi}^{2}}{3}+\frac{1}{2}\sqrt{\frac{8}{3}}H^{2}\beta\dot{\varphi}^{3}
\end{equation}
\begin{equation}
C=1-\frac{4H^{2}\beta\dot{\varphi}^{2}}{3},
\end{equation}
\begin{equation}
D=-\frac{24H^{2}\beta\dot{\varphi}^{2}}{3}\Big(\dot{a}+a\frac{\dot{H}}{H}
+\frac{\ddot{\varphi}}{\dot{\varphi}}\Big),
\end{equation}
\begin{equation}
E=\frac{1}{2}\Bigg(\dot{\varphi}^{2}-\frac{24H^{2}\beta\dot{\varphi}^{2}}{3}\Big[1+2\dot{a}+2a\frac{\dot{H}}{H}\Big]
+2\sqrt{\frac{8}{3}}H^{2}\beta\dot{\varphi}^{3}\Big[1+\dot{a}+a\frac{H}{H}+\frac{\ddot{\varphi}}{\dot{\varphi}}
\Big]\Bigg).\\
\end{equation}
We also include the $(0j)$ component of the gravitation field
equations in the same limit
\begin{equation}
\frac{3}{2}\,a^{2}H^{2}\theta\,\Omega_{m}=(B-A)\Delta\Psi+F\frac{\Delta\delta\varphi}{\dot{\varphi}}
-A\Delta\dot{\Upsilon}-B\frac{\Delta\dot{\delta\varphi}}{\dot{\varphi}}.
\end{equation}
The energy-momentum conservation equations are [14]
\begin{equation}
\dot{\delta}=-\theta
\end{equation}
and
\begin{equation}
\dot{\theta}+\theta+(\dot{a}+a\frac{\dot{H}}{H})\theta=-\frac{\Delta\Psi}{a^{2}H^{2}}
\end{equation}
where $\delta\equiv\delta\rho/\rho$ is the matter density contrast
and $\theta$ is the divergence of the matter peculiar velocity
field. Finally, Poisson's equation in this setup is as follows
\begin{equation}
\Delta\Psi=\frac{3G_{eff}}{2G}\,a^{2}H^{2}\delta\Omega_{m}
\end{equation}
where $G_{eff}$, the effective gravitational constant, is
\begin{equation}
G_{eff}=G\frac{D^{2}+CE}{A^{2}E+2ABD-B^{2}C}\, .
\end{equation}
This equation ( which its derivation can be obtained in [14] with
more details) is the basis of our forthcoming arguments. After
calculation of the effective gravitational constant, in the next
section we construct a holographic dark energy model in the normal
branch of this DGP-inspired modified GBIG scenario. We note that
since the normal DGP branch is ghost-free and cosmologically stable,
there is no ghost instability in our modified GBIG scenario too.

\section{Holographic Dark Energy in Modified GBIG Scenario}
\subsection{General Formalism}
To study holographic dark energy model in the modified GBIG
scenario, we first present a brief overview of the holographic dark
energy model [6]. It is well-known that the mass of a spherical and
uncharged D-dimensional black hole is related to its Schwarzschild
radius by [15]
\begin{equation}
M_{BH}=r_{s}^{D-3}\Big(\sqrt{\pi}M_{D}\Big)^{D-3}M_{D}
\frac{D-2}{8\Gamma\Big(\frac{D-1}{2}\Big)}
\end{equation}
where the D-dimensional Planck mass, $M_{D}$, is related to the
D-dimensional gravitational constant $G_{D}$ and the usual
4-dimensional Planck mass through $ M_{D}=G_{D}^{-\frac{1}{D-2}}$
and $M_{p}^{2}=M_{D}^{D-2}V_{D-4}$ with $V_{D-4}$ the volume of the
extra-dimensional space. If $\rho_{\Lambda D}$ is the bulk vacuum
energy, then application of the holographic dark energy proposal in
the bulk gives
\begin{equation}
\rho_{\Lambda D} {\cal{V}}(S^{D-2})\leq
r_{D-3}\Big(\sqrt{\pi}M_{D}\Big)^{D-3}M_{D}
\frac{D-2}{8\Gamma\Big(\frac{D-1}{2}\Big)}
\end{equation}
where ${\cal{V}}(S^{D-2})$ is the volume of the maximal hypersphere
in a D-dimensional spacetime given by
${\cal{V}}(S^{D-2})=A_{D}r^{D-1}$. $A_{D}$ is defined as
$$A_{D}=\frac{\pi^{\frac{D-1}{2}}}{\Big(\frac{D-1}{2}\Big)!}$$
$$A_{D}=\frac{\Big(\frac{D-2}{2}\Big)!}{(D-1)!}2^{D-1}\pi^{\frac{D-2}{2}}$$
for $D-1$ being even or odd respectively. By saturating inequality
(27), introducing $L$ as a suitable large distance ( IR cutoff) and
$c^{2}$ as a numerical factor, the corresponding vacuum energy as a
holographic dark energy is given by [15]
\begin{equation}
\rho_{\Lambda D}=c^{2}\Big(\sqrt{\pi}M_{D}\Big)^{D-3}M_{D}A_{D}^{-1}
\frac{D-2}{8\Gamma\Big(\frac{D-1}{2}\Big)}L^{-2}.
\end{equation}
Using this expression, one can calculate the corresponding pressure
via continuity equation and then the equation of state parameter of
the holographic dark energy defined as
$\omega_{\Lambda}=\frac{\rho_{\Lambda}}{p_{\Lambda}}$ can be
obtained directly.

Now we use this formalism in our GBIG setup. There are alternative
possibilities to choose $L$ ( see [15]). Here and in our forthcoming
arguments we choose the IR cut-off,\, $L$ \,, to be the crossover
scale $r$ which is related to the present Hubble radius via $r=1.26
H_{0}^{-1}$ ( see for instance the paper by Lue in Ref. [5] and
[17]) where $H$ is given by (15). In this case the effective
holographic dark energy density is defined as follows
\begin{equation}
\rho_{eff}=\frac{3H^{2}}{8\pi G_{eff}}
\end{equation}
where $H$ and $G_{eff}$ are given by (15) and (30) respectively. So
we find
\begin{equation}
\rho_{eff}=\frac{3H^{2}}{8\pi
G}\Big(\frac{A^{2}E+2ABD-B^{2}C}{D^{2}+CE}\Big).
\end{equation}\\
Using the conservation equation
\begin{equation}
\dot{\rho}_{eff}+3H(\rho_{eff}+p_{eff})=0,
\end{equation}
we can deduce
$$p_{eff}=H(D^{2}+CE)^{-2}\bigg[(2D\dot{D}+\dot{C}E+C\dot{E})(A^{2}E+2ABD-B^{2}C)-
(D^{2}+CE)(2A\dot{A}E+A^{2}\dot{E}$$
\begin{equation}
+2\dot{A}BD+2A\dot{B}D+2AB\dot{D}-2B\dot{B}C-B^{2}C)\bigg]-\Big(\frac{2\dot{H}+3H^{2}}{8\pi
G}\Big)\Big(\frac{A^{2}E+2ABD-B^{2}C}{D^{2}+CE}\Big).
\end{equation}
Hence, the equation of state parameter of the model is given as
follows
\begin{tiny}
\begin{equation}
\omega_{eff}=\frac{1}{3H}\frac{(2D\dot{D}+\dot{C}E+C\dot{E})(A^{2}E+2ABD-B^{2}C)-
(D^{2}+CE)(2A\dot{A}E+A^{2}\dot{E}+2\dot{A}BD+2A\dot{B}D
+2AB\dot{D}-2B\dot{B}C-B^{2}C)}{(A^{2}E+2ABD-B^{2}C)(D^{2}+CE)}-\frac{2\dot{H}+3H^{2}}{3H^{2}}.
\end{equation}
\end{tiny}
Now to proceed further, we should specify the form of $f(R)$. In
which follows we present a relatively general example by adopting
the Hu-Sawicki model.

\subsection{ An Example}
As an example, we consider the following observationally suitable
form of $f(R)$ ( the Hu-Sawicki model [16])
\begin{equation}
f(R)=R-R_{c}\frac{\vartheta(R/R_{c})^{m}}{1+(R/R_{c})^{m}},
\end{equation}
where for an spatially flat FRW type universe,
\begin{equation}
R=-6(\dot{H}+2H^{2})
\end{equation}
and both  $\vartheta$ and $R_{c}$ are free positive parameters. We
also adopt the ansatz with  $a=a_{0}t^{\nu}$ and
$\varphi=\varphi_{0}t^{-\mu}$ (with $\nu$ and $\mu$ positive
constants) and $\alpha(\varphi)=\frac{1}{2}(1-\xi\varphi^{2})$ with
$\xi$ a positive non-minimal coupling. We set
$V(\varphi)=V_{0}\varphi^{2}$ and we translate all of our
cosmological dynamics equations in terms of the redshift using the
relation between $a$ and $z$ as $1+z=\frac{a_{0}}{a}$. One can use
these ansatz in the Friedmann, Klein-Gordon and the conservation
equations to find constraints on the values of $\nu$ and $\mu$ ( and
also $\xi$) to have an accelerating phase of expansion at late-time
( to see a typical analysis in this direction see [17]). Based on
such an analysis, in which follows we set $\nu=1.2$ and $\mu=0.5$.
The value adopted for $\nu$ gives a late-time accelerating universe.
Then we perform some numerical analysis of the model parameter
space. We set $\xi=0.15$ which is close to the conformal coupling.
The reason for adopting such a value of $\xi$ is motivated from
constraint on $\xi$ from recent observations [18]. For our numerical
purposes we have set $V_{0}=1$ and $\beta=0.33$. Figure $1$ shows
the behavior of the effective energy density, $\rho_{eff}$, versus
the redshift. We see that the effective energy density grows with
time and always $\rho_{eff}>0$. This is a typical behavior of
phantom-like dark energy density. Figure $2$ shows the behavior of
$1+\omega_{eff}$ versus the redshift. With the choice of the model
parameters as we have adopted here, the universe enters the phantom
phase at $z\simeq0.25$ close to the observationally supported value.
\begin{figure}[htp]
\begin{center}\includegraphics{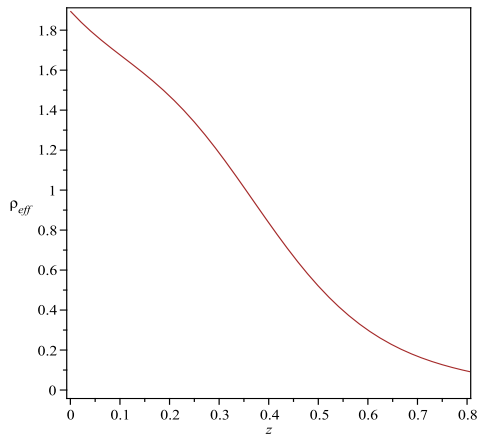} \vspace{6.5cm}
\end{center}
 \caption{\small { Variation of the effective dark energy density versus the redshift. }}
\end{figure}

\begin{figure}[htp]
\begin{center}\includegraphics{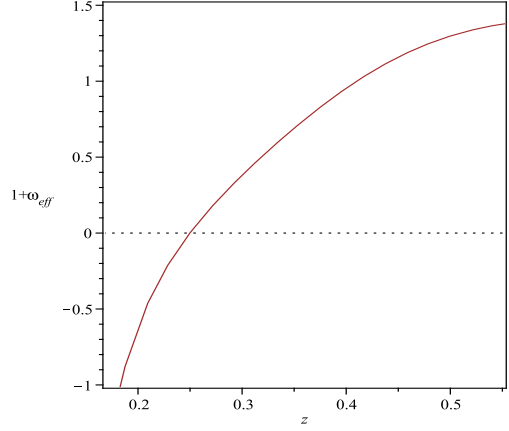} \vspace{5cm}
\end{center}
 \caption{\small { Variation of $1+\omega_{eff}$ versus the redshift. }}
\end{figure}
The deceleration parameter defined as
\begin{equation}
q=-\Big(\frac{\dot{H}}{H^{2}}+1\Big),
\end{equation}
in our setup takes the following form
\begin{tiny}
\begin{equation}
q=-\frac{8\beta
\big(\dot{f}(R)\alpha(\varphi)+\dot{\alpha}(\varphi)f(R)\big)}{rf^{2}(R)\alpha^{2}(\varphi)}
\Bigg[2\sqrt{1-3b}\,\cosh(\frac{\eta}{3})-1\Bigg]^{-1}-\frac{8\beta}{rf(R)\alpha(\varphi)}
\Bigg[2\sqrt{1-3b}\,\cosh{\frac{\eta}{3}}-1\Bigg]^{-2}\Bigg\{\frac{-3\dot{b}}{\sqrt{1-3b}}\cosh{\frac{\eta}{3}}
+\frac{2}{3}\sqrt{1-3b}\dot{\eta}\sinh{\frac{\eta}{3}}-1\Bigg\}-1
\end{equation}
\end{tiny}
Figure $3$ shows the behavior of $q$ versus $z$. In this model, the
universe has entered an accelerating phase in the past at
$z\simeq0.84$.

\begin{figure}[htp]
\begin{center}\includegraphics{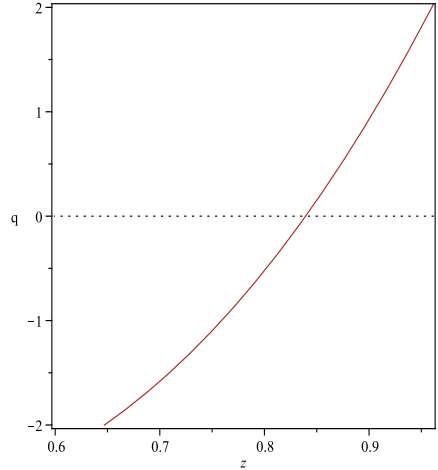} \vspace{6.5cm}
\end{center}
 \caption{\small { Variation of the deceleration parameter versus the redshift. }}
\end{figure}

Variation of $\dot{H}$ with cosmic time or redshift gives another
part of important information about the cosmology of this model. We
deduce the relation for variation of $H$ versus the cosmic time as
follows

\begin{tiny}
\begin{equation}
\dot{H}=\frac{r\big(\dot{f}(R)\alpha(\varphi)+f(R)\dot{\alpha}(\varphi)\big)}{8\beta}\Big[2\sqrt{1-3b}
\,\cosh(\frac{\eta}{3})
-1\Big]+\frac{rf(R)\alpha(\varphi)}{8\beta}\Big[\frac{-3\dot{b}}{\sqrt{1-3b}}\,\cosh(\frac{\eta}{3})
+\frac{2}{3}\sqrt{1-3b}\,\dot{\eta}\,\sinh(\frac{\eta}{3})\Big]
\end{equation}
\end{tiny}
Figure 4 shows the variation of $\dot{H}$ versus the redshift. Since
$\dot{H}<0$ always, the model universe described here will not
acquire super-acceleration and big-rip singularity in the future.

\begin{figure}[htp]
\begin{center}\includegraphics{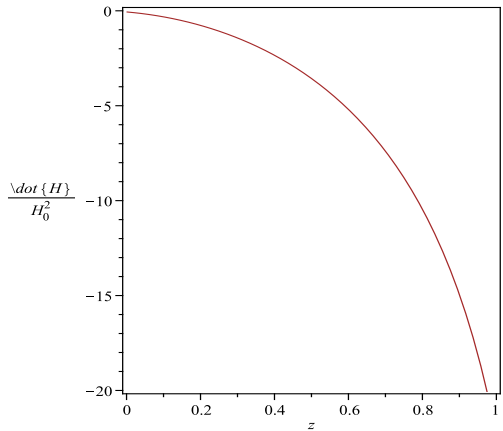} \vspace{5cm}
\end{center}
 \caption{\small { Variation of $\frac{\dot{H}}{H_{0}^{2}}$ versus the redshift. }}
\end{figure}

This example shows how a modified GBIG scenario has the potential to
give a phantom-like behavior in a holographic viewpoint. This
behavior is realized without introducing any phantom field in this
setup. In fact a combination of the curvature effect and the
non-minimal coupling provides the suitable framework for realization
of this phantom mimicry.
\section{Summary}
The model presented here, contains the Gauss-Bonnet term as the $UV$
sector of the theory, while the Induced Gravity effect completes the
$IR$ side of the model. The induced gravity on the brane is modified
in the spirit of $f(R)$ gravity which itself provides the facility
to self-accelerate even the normal, ghost-free branch of the
DGP-inspired model [11,19]. The model also considers a non-minimal
coupling between the canonical scalar field on the brane and the
modified induced gravity. This is a general framework for treating
dark energy problem and other alternative scenarios can be regarded
as subclasses of this general model. We have studied the
cosmological dynamics in this generalized braneworld setup within a
holographic point of view. By adopting a relatively general ansatz
for $f(R)$ gravity (Hu-Sawicki model) on the brane, we have shown
that the modified GBIG scenario presented here realizes the
phantom-like behavior: the effective energy density increases with
cosmic time and the effective equation of state parameter crosses
the phantom divide line smoothly in the same way as observations
suggest: from quintessence to the phantom phase. We have studied
also the holographic nature of the dark energy in this setup by
calculation of the effective gravitational constant via a
perturbational approach. In this model, $\dot{H}<0$ always, and
therefore the model universe described here will not acquire
super-acceleration and big-rip singularity in the
future.\\

{\bf Acknowledgment}\\
This work has been supported partially by Research Institute for
Astronomy and Astrophysics of Maragha, IRAN.

\end{document}